\documentstyle[prl,aps]{revtex}
\begin{document}
\draft
\title{Reply to:---Interlayer Josephson vortices in the high-$T_c$\\
superconducting cuprates}
\author{K. A. Moler}
\address{Department of Applied Physics, Stanford University, Stanford, CA
94305-4090}
\author{John R. Kirtley}
\address{IBM T.J. Watson Research Center, P.O. Box 218, Yorktown Heights,
NY 10598}
\maketitle
\vspace{0.5cm}
\baselineskip 18pt
\hsize 6.5 truein
\vsize 9.00 truein
\tolerance 10000

Farid raises the issue [1] of whether the Clem-Coffey solution [2,3]
is really appropriate to describe interlayer Josephson vortices in
layered superconductors.  We used this result to quantitatively
analyze our images of interlayer vortices in the high-temperature
layered cuprate superconductor Tl-2201 [4] in order to determine the
interlayer penetration depth, $\lambda_c$.  The length scales that appear in
this model are the interlayer spacing $s$, the in-plane penetration
depth $\lambda_a$, and $\lambda_c$.  For most cuprate superconductors, $a$
is a bit
over 10\ \AA, $\lambda_a$ is 0.1--0.2 microns, and $\lambda_c$ can be
microns and
depends strongly on the detailed chemical composition of the material, varying
greatly with small changes in oxygen doping.  In the usual description
of an interlayer Josephson vortex, the core extends a distance $\sim\!s$
perpendicular to the layers, and $\sim\!s\lambda_c/\lambda_a$ along the
layers [2,3]:
the field outside the core is described by the well-known anisotropic
London model.  For the sake of completeness in our paper, we fit our
data using the Coffey-Clem model, which includes an approximate
solution for the vortex core.  Because our experiment is only
sensitive to magnetic structure on micron length scales, the key
features of this model are, first, that the length scale for the
vortex core is less than a micron, and second, that the magnetic
fields outside the vortex core are described by the anisotropic London
model.  Any other model with these two characteristics would give the
same result for $\lambda_c$ within experimental error.

Farid points out the lack of an exact solution for the difficult
nonlinear problem of the structure of the vortex core, and implicitly
speculates that the correct solution may turn out to influence the
magnetic structure on length scales much larger than $s\lambda_c/\lambda_a$
and even
much larger than $\lambda_c$.  If this speculation is correct, our
interpretation that our images of interlayer Josephson vortices are a
direct measurement of the c-axis penetration depth, $\lambda_c \approx$ 20
microns in
optimally doped Tl-2201 [4], will be only one piece of a large body of
related experimental and theoretical work that will need to be
reevaluated.  We look forward to the opportunity to fit our data to a
theory including an exact treatment of both the vortex core and the
spreading associated with the superconductor-vacuum interface.

On the basis of related experimental evidence, it seems unlikely to us
that this exact theoretical solution will result in a qualitative
reevaluation of our results.  First, since our article [4] was
published, two independent groups using different optical techniques
have reported $\lambda_c$ = 17 microns [5] and $\lambda_c$ = 12 microns [6] in
optimally doped Tl-2201.  These results are both independent of the vortex
structure.  Second, we imaged vortices in the much-studied cuprate
superconductor LSCO, and found $\lambda_c\approx$ 5 microns [7].  This
unpublished
result is consistent with measurements by several other techniques
[8].  Therefore, if the exact theoretical solution indicates that the
size of an interlayer Josephson vortex is much larger than the
interlayer penetration depth, it will contradict the combined
experimental results of several groups on these two materials and will
require a new understanding of the optical as well as the magnetic
properties of layered superconductors.

It is a minor point that Farid misquotes our result as $\lambda_c\approx
22$ microns
in Tl-2201, when our article stated ``$\ldots$ we find $\lambda_c = 19 \pm
2$ microns.
There are larger systematic errors, which we estimate to be $<$30\%
$\ldots$''.
Incidentally, we have since studied what we view as the most likely
source of systematic error, namely the spreading of the vortex at the
superconductor-vacuum interface, and refined our estimate to $\lambda_c =
18 \pm
3$ microns [9].

Farid's observation that fitting our data to the Clem-Coffey model
does not give the correct values for $s$ and $\lambda_a$ should be taken as a
limitation of the data, and not necessarily the theory.  Our
measurements are made a few microns away from the sample surface.  The
detector is an 8 micron SQUID, fabricated with micron line widths and
shielded leads which may cause some distortion of the magnetic field
[10].  The data points are spaced every 1 micron, and the scan-to-scan
$x$-position is irreproducible on a 0.1-micron length scale due to the
tradeoffs required to get a large area scan.  It is unreasonable to
expect a sensible answer about any structure on the 0.001 micron
length scale or even the 0.1 micron length scale, no matter how exact
one's theoretical model.

Finally, Farid comments that the predicted relationship between $E^0_J$
and $\lambda_c$ [11], which we refuted through our measurement of
$\lambda_c$, may not
do ``justice to the ILT theory.''  We are grateful for a chance to
comment on this issue.  Based on a comparison of the published
predictions of one of the theory's authors [11] with the available
experimental data, ILT is not sufficient to explain the high critical
temperature and condensation energy in the cuprate superconductors
Tl-2201 [5] and Hg-1201 [12].  Some other mechanism must therefore be
in operation, perhaps in addition to the ILT mechanism.  In this
sense, ILT is not ``the'' theory of cuprate superconductivity.  The
interlayer tunneling model remains a creative and influential set of
ideas which may correctly describe some aspects of cuprate
superconductivity, as intriguing new evidence suggests [6,13].

We thank J. Berlinsky, J. R. Clem, and V. Kogan for useful
discussions.

\end{document}